\documentclass[a4paper,twoside,10pt]{article}
\usepackage{amsmath,amsthm,amssymb,mathrsfs}
\usepackage{bbm}
\usepackage[pdftex=true,colorlinks=true,linkcolor=blue,citecolor=red,bookmarks=true,bookmarksnumbered=true]{hyperref}
\usepackage[numbers,comma]{natbib}
\usepackage{fancyhdr}
\usepackage{titlesec}
\usepackage[a4paper,centering,bindingoffset=1cm,marginpar=2cm]{geometry}
\usepackage[font=normalsize,format=plain,labelfont=sc,textfont=up,width=0.75\textwidth,labelsep=period]{caption}
\titleformat{\section}{\filcenter\sc\large}{\thesection.\;}{0em}{}
\titleformat{\subsection}[runin]{\bf}{\thesubsection.\;}{0em}{}[.]

\hypersetup
{
    pdfauthor={Peter Elbau, Otmar Scherzer},
    pdfsubject={Quantitative Photoacoustic Imaging},
    pdftitle={Reconstruction Formulas for a Single Scattering Model in Photoacoustic Imaging},
    pdfkeywords={Quantitative Photoacoustic Imaging, Sectional Imaging, Reconstruction Formulas, Single Scattering}
}

\newcommand{\N}{\mathbbm{N}}
\newcommand{\R}{\mathbbm{R}}

\renewcommand{\d}{\,\mathrm d}

\DeclareMathOperator{\Div}{div}
\DeclareMathOperator{\tr}{tr}

\title{Reconstruction Formulas for a Single Scattering Model in Photoacoustic Imaging and Applications to Sectional Imaging}

\author{P. Elbau\and O. Scherzer\thanks{The work has been supported by the Austrian Science Fund (FWF)
within the national research network Photo\-acoustic Imaging in Biology and Medicine,
project S10505-N20.}}

\begin{document}
\maketitle

\begin{abstract}
There has been devoted significant mathematical research to model the light propagation in tissue and to recover the absorption and scattering coefficients after photoacoustic inversion. Typically, the basic light propagation models considered there are the diffusion limits of the radiative heat transfer model. These equations are well suited for models where the elastic scattering is the dominant effect. If, however, the scattering is less pronounced, a single scattering approximation model is practicable. As we show in this paper, this approach is practically relevant in focused/sectional photoacoustic imaging. In this paper, we study analytical reconstruction formulas for the single scattering case. To realise the single scattering approach, we propose concrete physical experiments based on photoacoustical sectional imaging.
\end{abstract}

\section*{Introduction}
In photoacoustic imaging (see e.g.~\cite{XuWan06,LiWan09,KucKun08,Wan08} for some mathematical and physical review papers), the interior of a small object is analysed by illuminating it with a short laser pulse and observing the acoustic wave which is hereby induced via the photoacoustic effect. The measurement of this pressure wave allows us (usually under the assumption that the acoustic wave travels with constant velocity and is free from any attenuation effects) to recover internal measurements in the form of the initially generated pressure
\[ P^{(0)}(x) = \gamma(x)\mu_{\mathrm a}(x)\bar\Phi(x),\quad x\in\R^3, \]
where $\gamma$ denotes the Gr\"uneisen parameter, which describes the change in pressure as energy is absorbed, $\mu_{\mathrm a}$ is the optical absorption coefficient of the material, and $\bar\Phi$ is the light fluence. Often, the variations in $\gamma$ and $\bar\Phi$ are neglected, so that $P^{(0)}$ can be considered to be proportional to the absorption coefficient, which is used to characterise the material, see~\cite{FinPatRak04,XuWan05} for some common reconstruction formulas for the initial pressure $P^{(0)}$.

Recent attempts have been made to model the light propagation in the material and to recover, so to speak in a second step, from the internal measurements $P^{(0)}$ for different illuminations $\bar\Phi$ the absorption coefficient $\mu_{\mathrm a}$ and the scattering coefficient, which enters in the light propagation model, see e.g.~\cite{Bal11b}. The basic light propagation models considered there are the diffusion limits of the radiative heat transfer model, the Boltzmann transport equation. These equations are well suited for models where the elastic scattering is the dominant effect. If, however, the scattering is less pronounced, a single scattering approximation model is practicable. This approach is practically relevant for focused/sectional photoacoustic imaging. In this paper, we study analytical reconstruction formulas for the single scattering case. Reconstruction formulas for the single scattering approach rely on a similar strategy as in the diffusion approximation \cite{Bal11b,Bal12} and are based on deriving equations for quotients of independent measurement data. To realise the single scattering approach, we propose concrete physical experiments based on photoacoustical sectional imaging.

\section{Light Propagation Models}
Considering only elastic scattering, the light propagation inside the object can be modelled with a Boltzmann transport equation, the so-called radiative transfer equation, for the density of photons $\psi_\vartheta(t,x)$ at the position $x\in\R^3$ at the time $t\in\R$ moving in the direction $\vartheta\in S^2$ of the form
\begin{equation}\label{eqBoltzmannEquation}
\frac1c\partial_t\psi_\vartheta(t,x)+\left<\vartheta,\nabla_x\psi_\vartheta(t,x)\right>+\mu_{\mathrm t}(x)\psi_\vartheta(t,x)=\frac{\mu_{\mathrm s}(x)}{4\pi}\int_{S^2}\Theta(\langle\tilde\vartheta,\vartheta\rangle)\psi_{\tilde\vartheta}(t,x)\d s(\tilde\vartheta),
\end{equation}
where the extinction or transport coefficent
\[ \mu_{\mathrm t}(x) = \mu_{\mathrm a}(x)+\mu_{\mathrm s}(x) \]
is given as the sum of the absorption coefficient $\mu_{\mathrm a}$ and the scattering coefficient $\mu_s$. Moreover, $c$ denotes the speed of light and $\Theta$ is the phase function, i.e.\ $\frac1{4\pi}\int_\Omega\Theta(\langle\tilde\vartheta,\vartheta\rangle)\d s(\tilde\vartheta)$ is the probability that a photon heading into a direction $\tilde\vartheta\in\Omega\subset S^2$ is scattered into the direction $\vartheta$.

In photoacoustic imaging, however, the excitation happens with a short laser pulse with some fixed frequency $\nu$, and we are not interested in the exact light distribution as a function of time, but only in the total energy being absorbed at each point. So, let us switch in a first step to the energy fluence $\Phi_\vartheta$ originating from photons moving in the direction $\vartheta\in S^2$ as new variable. We have the relation
\[ \Phi_\vartheta(x) = h\nu\int_{-\infty}^\infty\psi_\vartheta(t,x)c\d t,\quad x\in\R^3,\;\vartheta\in S^2, \]
where $h$ denotes the Planck constant.

\subsection*{Diffusion Approximation}
In quantitative photoacoustic tomography, see e.g.~\cite{Bal11b} for a review, one commonly uses the diffusion approximation of this transport equation which takes the form
\begin{equation}\label{eqDiffusionEquation}
\Div(\sigma\nabla\bar\Phi)(x)=\mu_{\mathrm a}(x)\bar\Phi(x),\quad\sigma(x)=\frac1{3(\mu_{\mathrm a}(x)+\mu_{\mathrm s}'(x))},\quad x\in\R^3,
\end{equation}
where
\begin{equation}\label{eqTotalLightFluence}
\bar\Phi(x) = \frac 1{4\pi}\int_{S^2}\Phi_\vartheta(x)\d s(\vartheta),\quad x\in\R^3,
\end{equation}
is the total light fluence at a point $x$ and $\mu'_{\mathrm s}$ is the so-called reduced scattering coefficient.

To obtain this approximation, the dependency from the direction $\vartheta$ is assumed to be at most linear. So we assume that the functions $\Phi_\vartheta$ and $\Theta$ can for all $x\in\R^3$ and all $\vartheta,\tilde\vartheta\in S^2$ approximatively be written as
\begin{equation}\label{eqLinearApproximation}
\Phi_\vartheta(x)\approx\phi_0(x)+\left<\vartheta,\phi_1(x)\right>\quad\text{and}\quad\Theta(\langle\tilde\vartheta,\vartheta\rangle)\approx\Theta_0+\Theta_1\langle\tilde\vartheta,\vartheta\rangle
\end{equation}
for some (sufficiently smooth) functions $\phi_0:\R^3\to\R$, $\phi_1:\R^3\to\R^3$, and some constants $\Theta_0,\Theta_1\in\R$.
Then we find for all $x\in\R^3$ and all $\vartheta\in S^2$ that
\[ \left<\vartheta,\nabla_x\Phi_\vartheta(x)\right> \approx \left<\vartheta,\nabla\phi_0(x)\right>+\sum_{j,k=1}^3\vartheta_j\vartheta_k\partial_{x_j}\phi_{1,k}(x) = \left<\vartheta,\nabla\phi_0(x)\right>+\vartheta^{\mathrm T}\mathrm D\phi_1(x)\vartheta \]
and
\[ \int_{S^2}\Theta(\langle\tilde\vartheta,\vartheta\rangle)\Phi_{\tilde\vartheta}(x)\d s(\tilde\vartheta)\approx 4\pi\Theta_0\phi_0(x)+\Theta_1\int_{S^2}\langle\tilde\vartheta,\vartheta\rangle\langle\tilde\vartheta,\phi_1(x)\rangle\d s(\tilde\vartheta), \]
where we used that the linear terms in $\tilde\vartheta$ give zero when integrated over $S^2$. To evaluate the integral therein, we switch to spherical coordinates with the polar angle $\alpha$ chosen as $\cos\alpha=\langle\tilde\vartheta,\vartheta\rangle$ and get for all $\vartheta\in S^2$, since the components orthogonal to $\vartheta$ vanish due to the axial symmetry of the integrand, that

\[ \int_{S^2}\langle\tilde\vartheta,\vartheta\rangle\tilde\vartheta\d s(\tilde\vartheta) =
   \int_{S^2}\langle\tilde\vartheta,\vartheta\rangle^2\vartheta\d s(\tilde\vartheta) =
   \left(2\pi\int_0^\pi\cos^2\alpha\sin\alpha\d\alpha\right)\vartheta = \frac{4\pi}3\vartheta. \]

So, plugging the approximation~\eqref{eqLinearApproximation} into the transport equation~\eqref{eqBoltzmannEquation}, which we average over time, we find for all $x\in\R^3$ that
\begin{multline}\label{eqMultipolExpansion}
\tfrac13\tr(\mathrm D\phi_1(x))+(\mu_{\mathrm a}(x)+\mu_{\mathrm s}(x))\phi_0(x)-\mu_{\mathrm s}(x)\Theta_0\phi_0(x) \\
+\left<\vartheta,\nabla\phi_0(x)+(\mu_{\mathrm a}(x)+\mu_{\mathrm s}(x))\phi_1(x)-\tfrac13\mu_{\mathrm s}\Theta_1\phi_1(x)\right>\\
+\vartheta^{\mathrm T}\left(\mathrm D\phi_1(x)-\tfrac13\tr(\mathrm D\phi_1(x))\right)\vartheta \approx 0.
\end{multline}
Ignoring the last term as it is of second order in $\vartheta$ (since $\vartheta^{\mathrm T}\vartheta=1$, we had to split off the trace term as it is of zeroth order) and using that
\[ \Theta_0\approx\frac1{4\pi}\int_{S^2}\Theta(\langle\tilde\vartheta,\vartheta\rangle)\d s(\tilde\vartheta)=1\quad\text{and}\quad\phi_0(x)\approx\frac1{4\pi}\int_{S^2}\Phi_\vartheta(x)\d s(\vartheta)=\bar\Phi(x),\quad x\in\R^3, \]
we get by equating the coefficients in~\eqref{eqMultipolExpansion} the equation system
\begin{align*}
\tfrac13\Div\phi_1(x) + \mu_{\mathrm a}(x)\bar\Phi(x) &\approx 0,\quad x\in\R^3, \\
\nabla\bar\Phi(x)+(\mu_{\mathrm a}(x)+\mu_{\mathrm s}'(x))\phi_1(x) &\approx0,\quad x\in\R^3,
\end{align*}
where $\mu_{\mathrm s}'(x) = (1-\frac13\Theta_1)\mu_{\mathrm s}(x)$ is the reduced scattering coefficient in $x\in\R^3$. Plugging $\phi_1$ from the second equation in the first one, we arrive at the diffusion equation~\eqref{eqDiffusionEquation}. A more general derivation for this diffusion equation, considering also higher order multipole expansions of $\Phi_\vartheta$ and $\Theta$, can be found in~\cite{Arr99}.

\subsection*{Single Scattering Approximation}
This diffusion model is well suited for materials where elastic scattering is the dominant effect. If, in contrast, the optical scattering is less pronounced, we may consider a single scattering model. In this case, we assume that scattered photons have a very low probability to interact again with the object, so that we can simply treat them as lost and not track their movement anymore.
We thus neglect the scattering term on the right hand side of the transport equation~\eqref{eqBoltzmannEquation} and get for the light fluence $\Phi_\vartheta$ from photons moving in the direction $\vartheta\in S^2$ the simplified transport equation
\begin{equation}\label{eqSingleScatteringEquation}
\left<\vartheta,\nabla_x\Phi_\vartheta(x)\right>=-\mu_{\mathrm t}(x)\Phi_\vartheta(x),\quad x\in\R^3.
\end{equation}

To specify boundary conditions for this differential equation, let us assume that no absorption or scattering occurs outside the object. Then the light fluence $\Phi_\vartheta$ is outside the object constant along all lines parallel to $\vartheta$. Thinking of the light fluence $\Phi_\vartheta$ being generated by a laser, which we for simplicity place infinitely far away in the direction $-\vartheta\in S^2$, the fluence between the laser and the object should be known from the specifications of the laser.
This means that we have for every $\vartheta\in S^2$ that
\begin{equation}\label{eqInitialConditionSingleScattering}
\lim_{\lambda\to-\infty}\Phi_\vartheta(y+\lambda\vartheta)=\Phi^{(0)}_\vartheta(y),\quad y\in E_\vartheta,
\end{equation}
for some known inital light fluence $\Phi^{(0)}_\vartheta:E_\vartheta\to[0,\infty)$ where
\[ E_\vartheta = \{y\in\R^3\mid \left<y,\vartheta\right>=0\}. \]

To formulate the solution of the differential equation~\eqref{eqSingleScatteringEquation} in a compact form, let us introduce for $\vartheta\in S^2$ the notation
\begin{equation}\label{eqCoordinateDecompositon}
x=x_\vartheta^\perp+x_\vartheta\vartheta,\quad x_\vartheta^\perp\in E_\vartheta,\;x_\vartheta\in\R,
\end{equation}
for the decomposition of a space coordinate $x\in\R^3$ into a vector $x_\vartheta^\perp\in\R^3$ orthogonal to $\vartheta$ and the component $x_\vartheta\vartheta=\left<x,\vartheta\right>\vartheta$ in direction of $\vartheta$.
With the boundary data~\eqref{eqInitialConditionSingleScattering}, the solution of the transport equation~\eqref{eqSingleScatteringEquation} is then given by
\[ \Phi_\vartheta(x) = \Phi^{(0)}_\vartheta(x_\vartheta^\perp)\exp\left(-\int_{-\infty}^{x_\vartheta}\mu_{\mathrm t}(x_\vartheta^\perp+\lambda\vartheta)\d\lambda\right) \]
for all $x\in\R^3$ and all $\vartheta\in S^2$.

In particular, if we illuminate the object with just one laser beam of photons moving in a direction $\vartheta_0\in S^2$, we have an initial light fluence of the form
\begin{equation}\label{eqSingleScatteringSingleIllumination}
\Phi^{(0)}_\vartheta(x) = \bar\Phi^{(0)}(x_{\vartheta_0}^\perp)\delta_{\vartheta_0}(\vartheta),\quad x\in E_\vartheta,
\end{equation}
with the known initial total light fluence $\bar\Phi^{(0)}:E_{\vartheta_0}\to\R$ of the laser beam. Here, $\delta_{\vartheta_0}$ denotes the $\delta$-distribution on the sphere at the point~$\vartheta_0\in S^2$ defined by $\frac1{4\pi}\int_{S^2}f(\vartheta)\delta_{\vartheta_0}(\vartheta)\d s(\vartheta)=f(\vartheta_0)$ for all functions $f\in C^\infty(S^2)$. In this case, the resulting total light fluence $\bar\Phi$, defined in equation~\eqref{eqTotalLightFluence}, is given by
\begin{equation}\label{eqSingleScatteringLightFluence}
\bar\Phi(x) = \bar\Phi^{(0)}(x_{\vartheta_0}^\perp)\exp\left(-\int_{-\infty}^{x_{\vartheta_0}}\mu_{\mathrm t}(x_{\vartheta_0}^\perp+\lambda\vartheta_0)\d\lambda\right),\quad x\in\R^3.
\end{equation}

\section{Reconstruction Formulas}
Let us now consider a photoacoustic measurement. I.e.\ we illuminate an unknown object resulting in a total light fluence $\bar\Phi$. Then the photoacoustic effect generates an initial pressure\footnote{More precisely, we should refer to it as a pressure difference to the equilibrium pressure in the object.} $P^{(0)}$ which is proportional to the absorption coefficient $\mu_{\mathrm a}$ of the material and the total light fluence $\bar\Phi$ of the laser light:
\[ P^{(0)}(x) = \gamma(x)\mu_{\mathrm a}(x)\bar\Phi(x),\quad x\in\R^3. \]
The proportionality constant $\gamma$ is called the Gr\"uneisen parameter and describes the thermodynamic properties of the material.\footnote{We have $\gamma=\frac1{C_{\mathrm p}}\alpha c_{\mathrm s}^2$ where $\alpha$ denotes the thermal expansion coefficient, $c_{\mathrm s}$ is the speed of sound, and $C_{\mathrm p}$ is the specific heat capacity at constant pressure.} See e.g.\ \cite{CoxLauBea09} for a derivation of this relation.

This initial pressure then initiates an acoustic wave propagating through the object. Usually, the simplified model of a homogeneous, elastic medium with constant speed of sound $c_{\mathrm s}$ is made, which leads to the linear wave equation
\begin{alignat}{2}
\partial_{tt}P(t,x)&=c_{\mathrm s}^2\Delta_xP(t,x),\quad &&t>0,\;x\in\R^3, \nonumber \\
\partial_t P(0,x)&=0,&&x\in\R^3, \label{eqWaveEquation} \\
P(0,x)&=P^{(0)}(x),&&x\in\R^3, \nonumber
\end{alignat}
for the pressure $P(t,x)$ at a point $x\in\R^3$ at time $t>0$.

There are a lot of articles discussing the problem of how to recover the initial pressure $P^{(0)}$ from some photoacoustic measurements $m$ of the acoustic wave $P$ outside the object.
For classical measurements where the function $m$ is of the form $m:[0,\infty)\times\partial X\to\R$ with $m(t,x)=P(t,x)$ for some domain $X$ containing the object, explicit reconstruction formulas could be obtained by Fourier methods~\cite{XuWan02a,XuFenWan02,XuXuWan02} leading to the so-called universal back-projection formula~\cite{XuWan05,Nat11}, which holds at least for the case where $X$ is a half-plane, a cylinder, or an ellipsoid. Other approaches~\cite{FinPatRak04,FinHalRak07,Kun07,Kun07b,Kun11a}, reducing the problem to the inversion of the spherical means operator, also led to explicit reconstruction formulas for simple geometries~$X$. And also from the more general setting of integral geometry used in~\cite{Pal11,Pal11b}, reconstruction formulas for the photacoustic problem could be obtained.

In the context of this paper, we simply want to assume that we are in some way able to recover from the photoacoustic measurements of the acoustic wave $P$ outside the object the initial pressure $P^{(0)}$ and thus have given the internal data
\begin{equation}\label{eqInitialPressures}
P^{(0)}_i(x)=\gamma(x)\mu_{\mathrm a}(x)\bar\Phi_i(x),\quad x\in\R^3,\;i=1,\ldots,N,
\end{equation}
for a certain number $N\in\N$ of different illuminations of the object resulting in different total light fluences $\bar\Phi_i$. The aim would be to recover from these data the three material parameters: the Gr\"uneisen parameter $\gamma$, the absorption coefficient $\mu_{\mathrm a}$, and the diffusion coefficient $\sigma$ (given by~\eqref{eqDiffusionEquation}) or the scattering coefficient $\mu_{\mathrm s}$, depending on the choice of light scattering model. However, as it turns out, only two of these three parameters can be recovered as a function of the third from this sort of data.

\subsection*{Diffusion Model}
 The focus of this work are explicit reconstruction methods for the single scattering approach. In comparison, the reconstruction for the parameters of the diffusion model~\eqref{eqDiffusionEquation} are discussed in the papers~\cite{AmmCapKanKoz09,CapFehGouKav09,Bal11b,Bal12}. To give an idea on what these methods are based, we shortly review the approach presented in~\cite{Bal12}.

The basis for explicit reconstructions there, as well as in our approach for the single scattering approximation, is to consider quotients of the known initial pressure data $P^{(0)}_i$, given by~\eqref{eqInitialPressures}, for multiple illuminations $i=1,\ldots,N$. Indeed, from the diffusion equation~\eqref{eqDiffusionEquation}, it follows (provided that $P^{(0)}_N(x)\ne0$ for all $x\in\R^3$) that the quotients
\[ u_i(x)=\frac{P^{(0)}_i(x)}{P^{(0)}_N(x)}=\frac{\bar\Phi_i(x)}{\bar\Phi_N(x)},\quad x\in\R^3,\; i=1,\ldots N-1, \]
fulfil for all $x\in\R^3$ and all $i\in\{1,\ldots,N-1\}$ the equation
\[ \Div(\sigma\bar\Phi_N^2\nabla u_i)(x) = 0,\quad\text{i.e.}\quad \left<\frac{\nabla(\sigma\bar\Phi_N^2)(x)}{\sigma(x)\bar\Phi_N^2(x)},\nabla u_i(x)\right>=-\Delta u_i(x). \]
So, if at each point $x\in\R^3$, the vectors $\nabla u_i(x)$, $i=1,\ldots,N-1$, span the whole space, we explicitly get the function $\nabla\log(\sigma\bar\Phi_N^2)$ and can thus recover from some initial conditions (in principle from the knowledge of $\sigma\bar\Phi_N^2$ at one point) the function $v$ given by
\[ v(x) = \sqrt{\sigma(x)}\bar\Phi_N(x),\quad x\in\R^3. \]
From this function, we can then get the two combinations
\[ \frac{P^{(0)}_N}{v(x)}=\frac{\gamma(x)\mu_{\mathrm a}(x)}{\sqrt{\sigma(x)}}\quad\text{and}\quad\frac{\Delta v(x)}{v(x)} = \frac{\mu_{\mathrm a}(x)}{\sigma(x)}+\frac{\Delta\sqrt\sigma(x)}{\sqrt{\sigma(x)}},\quad x\in\R^3, \]
of the three material parameters $\gamma$, $\mu_{\mathrm a}$, and $\sigma$. However, as it is shown in~\cite{BalRen11a}, for given Dirichlet boundary data for all the physical parameters, these two combinations already uniquely determine the internal data $P^{(0)}_i$, so that no additional information can be extracted from the measurements. So, we can only express two of the three material properties as a function of the third, see again~\cite{BalRen11a}.

\subsection*{Single Scattering Model}
For the reconstruction formulas in the single scattering model, it is enough to consider the initial pressures $P^{(0)}_1$ and $P^{(0)}_2$ generated by two different illuminations to recover the material parameters. We choose to illuminate the object once with a laser beam from the direction $-\vartheta\in S^2$ and once with a laser beam from the opposite direction $\vartheta\in S^2$. According to formula~\eqref{eqSingleScatteringLightFluence} for the resulting total light fluences $\bar\Phi_1$ and $\bar\Phi_2$, respectively, we get with the notation~\eqref{eqCoordinateDecompositon} (using that $x_{-\vartheta}=-x_\vartheta$ and $x_{-\vartheta}^\perp=x_\vartheta^\perp$) for all $x\in\R^3$ the representations
\begin{align}
\bar\Phi_1(x) &= \bar\Phi^{(0)}_1(x_\vartheta^\perp)\exp\left(-\int_{-\infty}^{x_\vartheta}\mu_{\mathrm t}(x_\vartheta^\perp+\lambda\vartheta)\d\lambda\right)\quad\text{and}\label{eqFluence1} \\
\bar\Phi_2(x) &= \bar\Phi^{(0)}_2(x_\vartheta^\perp)\exp\left(-\int_{x_\vartheta}^\infty\mu_{\mathrm t}(x_\vartheta^\perp+\lambda\vartheta)\d\lambda\right)\label{eqFluence2}
\end{align}
with some known initial total light fluences $\bar\Phi^{(0)}_1,\bar\Phi^{(0)}_2:E_\vartheta\to(0,\infty)$ (for simplicity, we assume as in the diffusion model before that the whole space is illuminated).
This we can plug into formula~\eqref{eqInitialPressures} to obtain the generated initial pressures
\begin{align}
P^{(0)}_1(x) &= \bar\Phi^{(0)}_1(x_\vartheta^\perp)\gamma(x)\mu_{\mathrm a}(x)\exp\left(-\int_{-\infty}^{x_\vartheta}\mu_{\mathrm t}(x_\vartheta^\perp+\lambda\vartheta)\d\lambda\right)\quad\text{and}\label{eqPressure1}\\
P^{(0)}_2(x) &= \bar\Phi^{(0)}_2(x_\vartheta^\perp)\gamma(x)\mu_{\mathrm a}(x)\exp\left(-\int_{x_\vartheta}^\infty\mu_{\mathrm t}(x_\vartheta^\perp+\lambda\vartheta)\d\lambda\right),\quad x\in\R^3.\label{eqPressure2}
\end{align}

Since only the combinations $\mu_{\mathrm t}=\mu_{\mathrm a}+\mu_{\mathrm s}$ and $\gamma\mu_{\mathrm a}$ of the three unknown parameters $\gamma$, $\mu_{\mathrm a}$, and $\mu_{\mathrm s}$ enter the expression for the initial pressure, there is no way to recover all three parameters, regardless of the number of measurements. Instead, we confine ourselves with reconstruction formulas for $\mu_{\mathrm t}$ and $\gamma\mu_{\mathrm a}$. As in the diffusion model, the additional knowledge of any of the three parameters then immediately allows the reconstruction of the two others (at least in the domain where the initial pressures do not vanish).

Let us denote with $\Omega$ the domain where the initial pressures do not vanish, i.e.\
\[ \Omega=\{x\in\R^3\mid P^{(0)}_1(x)P^{(0)}_2(x)>0\}. \]
Then for $x\in\Omega$, we find from~\eqref{eqPressure1} and~\eqref{eqPressure2} the identity
\[ \log\frac{\bar\Phi^{(0)}_1(x_\vartheta^\perp)P^{(0)}_2(x)}{\bar\Phi^{(0)}_2(x_\vartheta^\perp)P^{(0)}_1(x)}
=\int_{-\infty}^{x_\vartheta}\mu_{\mathrm t}(x_\vartheta^\perp+\lambda\vartheta)\d\lambda-\int_{x_\vartheta}^\infty\mu_{\mathrm t}(x_\vartheta^\perp+\lambda\vartheta)\d\lambda. \]
Thus, taking the derivative in the direction $\vartheta$, we end up with
\begin{equation}\label{eqFormulaForMuT}
\mu_{\mathrm t}(x)=\frac12\left<\vartheta,\nabla_x\log\frac{P^{(0)}_2(x)}{P^{(0)}_1(x)}\right>\quad\text{for all}\quad x\in\Omega.
\end{equation}
Assuming that
\begin{equation}\label{eqAssumptionOnScatteringCoefficient}
\mu_{\mathrm t}(x)=0\quad\text{for all}\quad x\in\R^3\setminus\Omega
\end{equation}
(meaning that whenever we have scattering or absorption at a point, this generates some pressure at that point), this completely determines the function $\mu_{\mathrm t}$. We can then get from this the product $\gamma\mu_{\mathrm a}$ with the relation
\begin{equation}\label{eqFormulaForGammaMu}
\gamma(x)\mu_{\mathrm a}(x) = \sqrt{\frac{P^{(0)}_1(x)P^{(0)}_2(x)}{\bar\Phi^{(0)}_1(x_\vartheta^\perp)\bar\Phi^{(0)}_2(x_\vartheta^\perp)}}\exp\left(\frac12\int_{-\infty}^\infty\mu_{\mathrm t}(x_\vartheta^\perp+\lambda\vartheta)\d\lambda\right),\quad x\in\R^3,
\end{equation}
which follows directly from the formulas~\eqref{eqPressure1} and~\eqref{eqPressure2}.

However, we remark that measuring additionally the total light fluences
\begin{equation}\label{eqLightFluenceBehindTheObject}
\bar\Phi_1^{(\infty)}(y)=\lim_{\lambda\to\infty}\bar\Phi_1(y+\lambda\vartheta)\quad\text{and}\quad\bar\Phi_2^{(\infty)}(y)=\lim_{\lambda\to\infty}\bar\Phi_2(y-\lambda\vartheta),\quad y\in E_\vartheta,
\end{equation}
of the laser light behind the object, the exponential factor in the formula~\eqref{eqFormulaForGammaMu} can be recovered from these measurements by
\[ \exp\left(\frac12\int_{-\infty}^\infty\mu_{\mathrm t}(y+\lambda\vartheta)\d\lambda\right) = \sqrt{\frac{\bar\Phi_i^{(0)}(y)}{\bar\Phi_i^{(\infty)}(y)}},\quad y\in E_\vartheta,\;i\in\{1,2\}, \]
which follows directly from~\eqref{eqFluence1} and~\eqref{eqFluence2}. With this additional measurement, we do no longer need to know the function $\mu_{\mathrm t}$ at all points to calculate $\gamma\mu_{\mathrm a}$ (in contrast to formula~\eqref{eqFormulaForGammaMu}). In particular, we do not require the assumption~\eqref{eqAssumptionOnScatteringCoefficient} to determine the function $\gamma\mu_{\mathrm a}$ via the formula
\begin{equation}\label{eqFormulaForGammaMu2}
\gamma(x)\mu_{\mathrm a}(x) = \sqrt{\frac{P^{(0)}_1(x)P^{(0)}_2(x)}{\bar\Phi^{(0)}_1(x_\vartheta^\perp)\bar\Phi^{(\infty)}_2(x_\vartheta^\perp)}} = \sqrt{\frac{P^{(0)}_1(x)P^{(0)}_2(x)}{\bar\Phi^{(\infty)}_1(x_\vartheta^\perp)\bar\Phi^{(0)}_2(x_\vartheta^\perp)}},\quad x\in\R^3.
\end{equation}

\section{Single Scattering Model in Photoacoustic Sectional Imaging}
A particular well suited example for the single scattering model is photoacoustic sectional imaging, see~\cite{GraPasNusPal11,NusGraPasPalMey12} and \cite{MaTarNtzRaz09,RazDisVinMaPerKoeNtz09} for some experimental results with photoacoustic sectional imaging and~\cite{ElbSchSchu12,KirSch11_report} for explicit reconstruction formulas for the initial pressure.

In this sectional setup, the object is not uniformly illuminated (as it is typically the case for standard photoacoustic imaging), but the laser light is focused so that only one slice of the object is illuminated. For simplicity, we want to assume that the illumination of the object is done with a single laser beam coming from a laser placed at infinity. As before, we consider the illumination from two opposite directions. And to further simplify the notation, we choose these directions to be $-\vartheta=(-1,0,0)$ and $\vartheta=(1,0,0)$. Moreover, this beam shall be perfectly focused onto the illumination plane $\{(x,y,z)\in\R^3\mid z=0\}$, so that the two corresponding initial total light fluences $\bar\Phi_1^{(0)}$ and $\bar\Phi_2^{(0)}$ of the laser beam, appearing in the equations~\eqref{eqFluence1} and~\eqref{eqFluence2}, have the form
\[ \bar\Phi_i^{(0)}(0,y,z) = \hat\Phi_i^{(0)}(y)\delta(z),\quad y,z\in\R,\;i\in\{1,2\}, \]
for some known functions $\hat\Phi_1^{(0)},\hat\Phi_2^{(0)}:\R\to(0,\infty)$.

In practice, of course, some scattering effects will still occur, leading to a larger illumination region inside the object. To diminish this effect, people started to use focusing detectors for the measurement of the acoustic waves, see e.g.~\cite{GraPasNusPal11,NusGraPasPalMey12}. These detectors are tuned in such a way that pressure waves originating from points outside of the desired illumination plane interfere destructively on the detector surface so that these waves contribute considerably less to the measurements than those originating from the illumination plane.

From the modelling point of view, these focusing detectors simply suppress the detection of those pressure waves which are generated by the absorption of scattered photons (unless they are only scattered inside the illumination plane or multiple times in a way that they end up being absorbed in the illumination plane again). On the other hand, this absorption of scattered photons is exactly the effect which we neglect in the single scattering model. Therefore, the single scattering model seems to be a good approximation for the modelling of photoacoustic sectional imaging with focusing detectors.

Remarking that for the direction $\vartheta=(1,0,0)$, the decomposition~\eqref{eqCoordinateDecompositon} simply reads $(x,y,z) = (0,y,z)+x(1,0,0)$, we get from the equations~\eqref{eqPressure1} and~\eqref{eqPressure2} that the resulting initial pressures $P^{(0)}_1$ and $P^{(0)}_2$ in this single scattering model are given by
\[ P^{(0)}_i(x,y,z) = \hat P^{(0)}_i(x,y)\delta(z),\quad x,y,z\in\R,\;i\in\{1,2\}, \]
where the functions $\hat P^{(0)}_1,\hat P^{(0)}_2:\R^2\to\R$ are defined by
\begin{align}
\hat P^{(0)}_1(x,y) &= \hat\Phi_1^{(0)}(y)\gamma(x,y,0)\mu_{\mathrm a}(x,y,0)\exp\left(-\int_{-\infty}^x\mu_{\mathrm t}(\lambda,y,0)\d\lambda\right)\quad\text{and}\label{eqSectionalPressure1}\\
\hat P^{(0)}_2(x,y) &= \hat\Phi_2^{(0)}(y)\gamma(x,y,0)\mu_{\mathrm a}(x,y,0)\exp\left(-\int_x^\infty\mu_{\mathrm t}(\lambda,y,0)\d\lambda\right),\quad(x,y)\in\R^2. \label{eqSectionalPressure2}
\end{align}

Modelling the propagation of the pressure wave with the linear wave equation~\eqref{eqWaveEquation}, we derived in~\cite{ElbSchSchu12} explicit reconstruction formulas for the initial pressures $\hat P^{(0)}_i$, $i\in\{1,2\}$, in the illumination plane for different detector setups, where the focusing detectors are approximated by either point, integrating line, or integrating plane detectors.

As the only difference between the expressions~\eqref{eqSectionalPressure1}, \eqref{eqSectionalPressure2} and~\eqref{eqPressure1}, \eqref{eqPressure2} is that the first ones do not depend on the distance $z$ to the illumination plane, we can use the exact same derivation as we used to get the formulas~\eqref{eqFormulaForMuT}, \eqref{eqFormulaForGammaMu}, and \eqref{eqFormulaForGammaMu2} to recover $\mu_{\mathrm t}$ and the product $\gamma\mu_{\mathrm a}$. We therefore find for $\mu_{\mathrm t}$ that
\[ \mu_{\mathrm t}(x,y,0) = \frac12\partial_x\log\frac{\hat P^{(0)}_2(x,y)}{\hat P^{(0)}_1(x,y)}\quad\text{for all}\quad (x,y)\in\hat\Omega, \]
where $\hat\Omega=\{(x,y)\in\R^2\mid \hat P^{(0)}_1(x,y)\hat P^{(0)}_2(x,y)>0\}$. And if we assume again that the transport coefficient $\mu_{\mathrm t}$ fulfils that $\mu_{\mathrm t}(x,y,0)=0$ if $(x,y)\in\R^2\setminus\hat\Omega$, then this allows us to calculate $\gamma\mu_{\mathrm a}$ via
\[ \gamma(x,y,0)\mu_{\mathrm a}(x,y,0) = \sqrt{\frac{\hat P^{(0)}_1(x,y)\hat P^{(0)}_2(x,y)}{\hat \Phi^{(0)}_1(y)\hat\Phi^{(0)}_2(y)}}\exp\left(\frac12\int_{-\infty}^\infty\mu_{\mathrm t}(\lambda,y,0)\d\lambda\right),\quad (x,y)\in\R^2. \]
As in the previous section, this can also be simplified to
\[ \gamma(x,y,0)\mu_{\mathrm a}(x,y,0)
= \sqrt{\frac{\hat P^{(0)}_1(x,y)\hat P^{(0)}_2(x,y)}{\hat\Phi^{(0)}_1(y)\hat\Phi^{(\infty)}_2(y)}}
= \sqrt{\frac{\hat P^{(0)}_1(x,y)\hat P^{(0)}_2(x,y)}{\hat\Phi^{(\infty)}_1(y)\hat\Phi^{(0)}_2(y)}},
\quad (x,y)\in\R^2, \]
with the additional measurement of the total light fluences
\[ \bar\Phi_i^{(\infty)}(0,y,z)=\hat\Phi_i^{(\infty)}(y)\delta(z),\quad y,z\in\R,\;i\in\{1,2\}, \]
behind the object as defined in~\eqref{eqLightFluenceBehindTheObject} with $\vartheta=(1,0,0)$.

\section*{Conclusion}
We have shown explicit reconstruction formulas for photoacoustic imaging for the three main physical parameters, the Gr\"uneisen parameter, the absorption and the scattering coefficient, in a single scattering light propagation model. Here, in analogy, what has been pointed out earlier \cite{Bal11b,Bal12} for the diffusion model, it is also only possible to recover two of them as a function of the third.

Moreover, we have argued that this single scattering model is a good approximation for photoacoustic sectional imaging where focusing detectors are used to measure only acoustic signals originating from the illuminated plane.

\def\cprime{$'$} \providecommand{\noopsort}[1]{}

\end{document}